\documentclass[namedreferences]{SolarPhysics}
\usepackage[optionalrh]{spr-sola-addons} 
\usepackage{graphicx}        
\usepackage{color}           
\usepackage{url}             
\usepackage{rotating}        
\usepackage{times}
\usepackage{multirow}
\begin{document}

\begin{article}

\begin{opening}

\title{A Statistical Study on the Morphology of Rays and Dynamics of Blobs in the Wake of Coronal Mass Ejections}

\author{H.Q. \surname{Song}$^{1,2}$\sep
        X.L. \surname{Kong}$^{1}$\sep
        Y. \surname{Chen}$^{1}$\sep
        B. \surname{Li}$^{1}$\sep
        G. \surname{Li}$^{1,3}$\sep
        S.W. \surname{Feng}$^{1}$\sep
        L.D. \surname{Xia}$^{1}$
       }
\runningauthor{H.Q. Song et al.} \runningtitle{A Statistical Study
on Dynamics of CME Blobs}

   \institute{$^{1}$ Shandong Provincial Key Laboratory of Optical Astronomy
   and Solar-Terrestrial Environment, School of Space Science and Physics,
   Shandong University at Weihai, Weihai, Shandong 264209,
                     China\\
                     email: \url{yaochen@sdu.edu.cn}\\
              $^{2}$ State Key Laboratory of Space Weather, Chinese Academy
              of Sciences, Beijing 100190, China\\
              $^{3}$ Department of Physics and CSPAR, University of Alabama in Huntsville, Huntsville, AL
              35899, USA
             }

\begin{abstract}
In this paper, with a survey through the Large Angle and
Spectrometric Coronagraph (LASCO) data from 1996 to 2009, we
present 11 events with plasma blobs flowing outwards sequentially
along a bright coronal ray in the wake of a coronal mass ejection.
The ray is believed to be associated with the current sheet
structure that formed as a result of solar eruption, and the blobs
are products of magnetic reconnection occurring along the current
sheet. The ray morphology and blob dynamics are investigated
statistically. It is found that the apparent angular widths of the
rays at a fixed time vary in a range of 2.1$^{\circ}$ --
6.6$^{\circ}$ (2.0$^{\circ}$ -- 4.4$^{\circ}$) with an average of
3.5$^{\circ}$ (2.9$^{\circ}$) at $3 R_\odot$ ($4 R_\odot$),
respectively, and the observed durations of the events vary from
12 h to a few days with an average of 27 h. It is also found,
based on the analysis of blob motions, that 58\% (26) of the blobs
were accelerated, 20\% (9) were decelerated, and 22\% (10) moved
with a nearly-constant speed. Comparing the dynamics of our blobs
and those that are observed above the tip of a helmet streamer, we
find that the speeds and accelerations of the blobs in these two
cases differ significantly. It is suggested that these differences
of the blob dynamics stem from the associated magnetic
reconnection involving different magnetic field configurations and
triggering processes.

\end{abstract}

\keywords{Coronal Mass Ejections, Electric Currents and Current
Sheets, Magnetic Reconnection}

\end{opening}

\section{Introduction}

A bright ray-like structure is frequently observed in the wake of
a coronal mass ejection (CME) observed with coronagraphs like
Large Angle and Spectrometric Coronagraph (LASCO) (Brueckner
\textit{et al.}, 1995) on the \textit{Solar and Heliospheric
Observatory} (SOHO) spacecraft (\textit{e.g.}, Ciaravella
\textit{et al.}, 2002; Vr\v snak \textit{et al.}, 2009;
Patsourakos and Vourlidas, 2011). Within the post-CME ray, a
current sheet structure seems to exist, connecting the post-CME
loops to the CME ejecta (\textit{e.g.}, Ciaravella \textit{et
al.}, 2002; Webb \textit{et al.}, 2003; Ko \textit{et al.}, 2003,;
Lin \textit{et al.}, 2005; Bemporad \textit{et al.}, 2006),
agreeing with theoretical predictions given by  both the classical
CSHKP flare model (Carmichael, 1964; Sturrock, 1966; Hirayama,
1974; Kopp and Pneuman, 1976) and some more recent CME models
(\textit{e.g.}, Lin and Forbes, 2000; Chen \textit{et al.}, 2007).

The existence of a current sheet within the ray is mainly
supported by the following three aspects of observations. First,
as already mentioned, for the earth-side eruptions with observable
flaring or post-CME loops, the ray seems to connect the top of the
loops to the bottom of the ejecta, which is consistent with
theoretical expectations. Second, it is found, through the
spectral analysis of the Ultraviolet Coronagraph Spectrometer
(UVCS) (Kohl, \textit{et al.}, 1995) data, that there exist
enhanced high-temperature emissions from ions like Fe {\sc xviii}
indicating that the plasma in the ray is much hotter than the
surrounding plasma (\textit{e.g.}, Ciaravella \textit{et al.},
2002; Bemporad \textit{et al.}, 2006; Ciaravella and Raymond,
2008). This agrees with the idea that the plasma is heated by
magnetic reconnection occurring along the assumed current sheet
structure. Third, fast-moving plasma blobs are sometimes observed
to move outwards along the ray, which are believed to be products
of magnetic reconnection, therefore supporting the presence of a
current sheet structure within the post-CME ray (Ko \textit{et
al.}, 2003; Lin \textit{et al.}, 2005).

However, the physical nature of the post-CME ray and its relation
to the current sheet structure is still controversial presently.
There exist at least three different scenarios. (1) It is
suggested that the ray as a whole represents a
significantly-broadened current sheet structure (\textit{e.g.}, Ko
\textit{et al.}, 2003; Lin \textit{et al.}, 2005). From
observations, it was seen that the ray can be as wide as tens to
hundreds of kilometers, which exceeds, by several orders of
magnitude, the width of a current sheet allowed by the classical
plasma theory (\textit{e.g.}, Litvinenko, 1996; Wood and Neukirch,
2005) under coronal conditions, which is only about tens of
meters. This gives rise to the most controversial aspect of this
scenario. Recently, Lin \textit{et al.} (2009) proposed some
broadening mechanisms mostly related to the turbulent reconnection
associated with tearing mode instability or a time-dependent
Petschek reconnection. (2) Liu \textit{et al.} (2009) suggested
that the post-CME ray may correspond to a plasma sheet structure
with an embedded current sheet, similar to the plasma--current
sheet configuration atop of a typical helmet streamer. In-situ
measurements show that the width of the heliospheric current sheet
is smaller than the associated plasma sheet by one to two orders
of magnitude (Winterhalter \textit{et al.}, 1994). Therefore, in
this scenario the requirement of the broadening mechanism can be
greatly reduced. However, the authors did not explain the
formation of the plasma sheet or the origin of the observed
high-temperature emission therein.

The above two scenarios both suggest that the current sheet is an
intrinsic part of the post-CME ray, which are distinctive from the
third one proposed by Vr\v snak \textit{et al.} (2009). They
suggest that the whole ray structure is a result of the exhaust of
a Petschek reconnection process with a diffusion region too low to
be observed by the present coronagraphs. According to this
scenario, the ray structure is constrained by a pair of slow mode
shocks, and does not contain the aforementioned type of current
sheet structure. We note that Lin \textit{et al.} (2009) has
pointed out a few difficulties, mainly associated with the
electron heating and acceleration, of using the Petschek
reconnection to account for the features observed during solar
eruptions and associated radio bursts.

Therefore, the physical nature of the post-CME ray and its
relationship with the current sheet structure remain elusive. On
the other hand, observations of plasma blobs flowing along the ray
shed more light on the ray properties as well as the relevant
reconnection process. So far, only a few CME-ray-blob events have
been studied in detail (Ko \textit{et al.}, 2003; Lin \textit{et
al.}, 2005). According to these studies, the plasma blobs are
products of magnetic reconnection along the current sheet which is
triggered mostly by the tearing mode instability. The initial blob
velocities can then be used to give the reconnection outflow
speed, or approximately the local Alfv\'en speed, which can be
used to further deduce the magnetic field strength in the current
sheet region with appropriate assumptions on densities.

For example, according to Ko \textit{et al.} (2003), the field
strength in the ray region for the event on 8 January 2002 is
determined to be 0.69 gauss (G) and 0.47 G at $3.1 R_\odot$ and
$4.5 R_\odot$, respectively. In Lin \textit{et al.} (2005), the
magnetic reconnection rate is estimated for the event on 18
November 2003 with the reconnection inflow speed determined by the
UVCS spectral observations. To our knowledge, there have been no
statistical studies published for the CME-ray-blob events. The
major purpose of this paper is to conduct such a study. The events
selected here are found via a survey through the LASCO
observations from 1996 to 2009. Both the morphology of the
post-CME rays and the blob dynamics are examined to understand the
rays as well as the associated reconnection process.

Similar to the above ray-blob phenomenon, there exists another
type of blobs flowing outwards along the bright rays atop of
helmet streamers (Sheeley, \textit{et al.,}1997; Wang \textit{et
al.}, 1998, 2000). In some of these events, the blobs are found to
be released in a persistent and quasi-periodic manner (Wang
\textit{et al.}, 2000; Song \textit{et al.}, 2009). This behavior
is modeled by Chen \textit{et al.} (2009) and interpreted as a
result of an intrinsic instability of coronal streamers exhibiting
a current-sheet and cusp morphology. The instability originates
from the magnetic topology associated with the streamer cusp, in
the neighborhood of which the magnetic field is too weak to
contain a plasma at the coronal temperature. Therefore, the plasma
expands outwards resulting in an elongation of the closed arcade
containing the streamer cusp. The elongation brings a
slowly-moving, confined plasma into the solar wind flowing along
the streamer stalk and introduces velocity shear into the system.
When the shear is large enough, a streaming sausage mode
instability (Lee \textit{et al.}, 1988; Wang \textit{et al.},
1988) may develop to trigger the reconnection, producing the
disconnected plasma blobs. Besides the sausage mode instability,
there are also two other possible processes contributing to the
occurrence of reconnection. First, the expansion of a closed
magnetic loop may produce thermal and magnetic pressure gradients
which cause the expanding loop to shrink and form a current sheet
or an X-type neutral point along the arcade (Wang \textit{et al.},
2000). Second, the effect of the converging flow of the solar wind
plasma may also play a role in the shrinkage of the expanding
field lines (\textit{e.g.}, Lapenta and Knoll, 2005).

In this paper, we will compare the dynamics of blobs moving along
the post-CME rays and those of the streamer blobs to provide more
clues on the associated magnetic reconnection process. The paper
is organized as follows: In Section 2, we will present the
observations and statistical results. We will then compare these
ray blobs with those observed in streamers in Section 3, and will
give our conclusions in Section 4.

\section{Observations and Results}

In the survey of the LASCO data from 1996 to 2009, we have
searched for solar eruptions with a bright and clear ray structure
in the wake of CME ejecta. We have imposed the criteria that at
least three blobs flow outwards along the rays. For all earth-side
events (i.e., except events 7 and 8) with observable post-CME
loops, we have also required that there exist associating post-CME
loops and the loops are generally co-aligned with the observed
rays. With these conditions, eleven events have been identified.
In Table 1, we first list the information of the CMEs and
associated flares if available. The first column is the sequence
number of the events. The second column gives the first appearance
time of the CME in the LASCO C2 field of view (FOV). The third
column is the central position angle (CPA) of the ejecta (as
measured counterclockwise from the north pole). The fourth column
is the linear speed of the CME front in units of km s$^{-1}$. The
next two columns list the starting time and location of the
associated flares. The symbol ``$\backslash$" means that no
associated flare was identified. The last column is the class of
the associated flares, defined by the soft X-ray flux measurements
of the \textit{Geostationary Operational Environmental Satellite}
(GOES). The CME parameters included in the table are adopted from
the database maintained by the CDAW data center
(http://cdaw.gsfc.nasa.gov/CME$\_$list/).

For four events (events 5, 9, 10, and 11), unambiguous flares were
identified. For the rest of the events, we were unable to identify
the accompanying flares. The reasons are the following. First,
events 7 and 8 occurred on the back side of the Sun and events 2,
3, and 6 occurred at the limb. The associated flares for these
events are therefore either totally or partially blocked by the
Sun. Second, for events 1, 2, 3, and 6, there were other events
occurring simultaneously at other locations of the Sun, making the
association of the detected flares with our CMEs very hard.
Finally, for event 4, the background X-ray emission was as strong
as the peak emission of a C3.0 flare; therefore, a possible
associated flare ($<$C3.0) can not be discerned.


\begin{table}[!htbp]
\begin{minipage}[t]{\columnwidth}
\caption{Information on the eleven CME events and associated
flares.
 }
\renewcommand{\thefootnote}{\alph{footnote}}
\renewcommand{\footnoterule}{}
\tabcolsep=3pt
\begin{tabular}{clllclcr}
  \hline
  No. & \multicolumn{3}{l}{CME}  & & \multicolumn{3}{l}{Associated flare} \\
  \cline{2-4}
  \cline{6-8}
    & First C2 appe- & CPA & Velocity  & & Starting & Location &
    Class
\\ & arance time (UT) &  & (km s$^{-1}$)& & time (UT) & &
\\
  \hline
1 & 1999/05/26 08:06 &321$^{\circ}$& 565      &   & $\backslash$      &    \\
2 & 2000/06/21 19:31 &239$^{\circ}$& 432\footnotemark[1]\footnotetext[1]{Limb events}     &   & $\backslash$     &  &  \\
3 & 2001/07/12 00:06 &240$^{\circ}$& 736\footnotemark[1]     &   & $\backslash$     &  & \\
4 & 2001/09/21 08:54 &135$^{\circ}$& 659      &   & $\backslash$      &    \\
5 & 2001/11/17 05:30 &Halo (OA)& 1379      &   & 04:49        & S18E42  &M2.8                \\
6 & 2002/01/08 17:54 &Halo (OA)& 1794\footnotemark[1]     &   & $\backslash$     &  & \\
7 & 2002/06/19 15:06 &58$^{\circ}$& 298       &   & $\backslash$ &     &   \\
8 & 2003/07/23 12:30 &213$^{\circ}$& 329      &   & $\backslash$ &    &    \\
9 & 2003/10/24 02:54 &113$^{\circ}$& 1055     &   & 02:27        & S19E72 &M7.6                \\
10& 2003/11/04 19:54 &Halo (OA)& 2657\footnotemark[1]    &   & 19:29        & S19W83&$>$X28              \\
11& 2003/11/18 09:50 &95$^{\circ}$& 1824\footnotemark[1]     &   & 09:23           & S13E89  &M4.5            \\
  \hline
\end{tabular}
\end{minipage}
\end{table}

Table 2 lists the morphological and dynamical parameters of the
post-CME rays and blobs in these events. The first column is the
sequence number of the events. The second column gives the
measurement time of the ray. The third column lists the direction
and CPA of the ray. The CPA is measured at the bottom of the C2
FOV, \textit{i.e.}, at about 2.2 $R_{\odot}$. The letter in the
third column represents the direction of the rays at the above
time with ``R'' meaning radial, ``P'' poleward, and ``E''
equatorward. The fourth column gives the angular widths of the
rays measured at $3 R_\odot$ and $4 R_\odot$, respectively, at the
time shown in the second column. The method for measuring the ray
width will be introduced later. The fifth column is the observed
time duration of the ray structure within the C2 FOV. The sixth
column is the number of blobs. The seventh column is the average
velocity range, and the last column is the acceleration range.

\begin{table}[!htbp]
\begin{minipage}[t]{\columnwidth}
\caption{Morphological and dynamical parameters of the post-CME
rays and blobs in the eleven events.
 }
\renewcommand{\footnoterule}{}
\tabcolsep=3pt
\begin{tabular}{cllccclcc}
 \hline
 No. & \multicolumn{4}{l}{Ray} & & \multicolumn{3}{l}{Blob} \\
 \cline{2-5}
 \cline{7-9}
& Observation  &  Direction & Angular &Duration & & Number &
Velocity & Acceleration
\\ & time (UT) & \& CPA &  width  & (h)& &
 & \footnote{The average velocities in this column are obtained by the
linear fit of all height vs. time points for each blob. } & Range
\\ & & & at 3 (4) $R_\odot$& & & &(km s$^{-1}$)&(m s$^{-2}$)
\\
  \hline
1 & 1999/05/26 11:26 &   P 327$^{\circ}$    & 3.9$^{\circ}$ (2.3$^{\circ}$) & 25 & &5 & 410--687 & -8.5--15.5\\
2 & 2000/06/22 05:30 &   R 230$^{\circ}$    & 5.1$^{\circ}$ (4.1$^{\circ}$) & 22 & &3 & 287--430 & 9.5--24.5\\
3 & 2001/07/12 16:30 &   E 222$^{\circ}$    & 3.5$^{\circ}$ (3.2$^{\circ}$) & 25 & &4 & 327--598 & -13.1--8.0\\
4 & 2001/09/21 18:30 &   P 116$^{\circ}$    & 3.0$^{\circ}$ (2.5$^{\circ}$) & 23 & &3 & 458--491 & -0.2--17.7\\
5 & 2001/11/17 14:06 &   P 77$^{\circ}$     & 2.8$^{\circ}$ (2.8$^{\circ}$) & 28 & &5 & 252--609 & -22.0--7.1\\
6 & 2002/01/09 15:54 &   P 82$^{\circ}$     & 2.4$^{\circ}$ (2.2$^{\circ}$) & 40 & &4 & 475--845 & -25.1--7.0\\
7 & 2002/06/20 09:30 &   E 57$^{\circ}$     & 3.8$^{\circ}$ (4.3$^{\circ}$) & 34 & &4 & 203--432 & -18.5--0.8\\
8 & 2003/07/24 05:30 &   E 220$^{\circ}$    & 6.6$^{\circ}$ (4.4$^{\circ}$) & 25 & &3 & 418--494 & -2.6--12.7\\
9 & 2003/10/24 06:54 &   P 100$^{\circ}$    & 2.6$^{\circ}$ (2.0$^{\circ}$) & 12 & &3 & 515--756 & 1.9--25.8\\
10& 2003/11/04 23:48 &   R 252$^{\circ}$    & 3.2$^{\circ}$ (2.4$^{\circ}$) & 25 & &5 & 586--786 & -26.1--22.9\\
11& 2003/11/18 14:50 &   E 97$^{\circ}$     & 2.1$^{\circ}$ (2.1$^{\circ}$) & 39 & &6 & 527--681 & 0.72--59.4\\
  \hline
\end{tabular}
\end{minipage}
\end{table}


It can be seen that most events took place in the years from 2001
to 2003, near the maximum of solar activity. The apparent linear
CME speeds vary in a large range from 298 to 2657 km s$^{-1}$. It
is found that in nine events the rays are non-radial with five
poleward and four equatorward. The difference in angles between
the CPA of the CME ejecta and the ray varies from 1$^\circ$ to
19$^\circ$ with an average of 9$^\circ$, indicating that the ray
direction is basically co-aligned with the centroid of the ejecta.
In addition, in events 6, 9, and 11 the ray PAs change apparently
by about 8$^\circ$ -- 10$^\circ$ during the observations, while
the changes are not discernible in the other eight events.

To measure the angular widths of the rays, we have applied a
Gaussian fitting to the white light brightness of rays at $3
R_\odot$ and $4 R_\odot$, and have adopted the full width at half
maximum (FWHM) as the ray width. It is found that the obtained ray
widths vary in a range of 2.1$^\circ$ -- 6.6$^\circ$ with an
average being 3.5$^\circ$ at $3 R_\odot$, and 2.0$^\circ$ --
4.4$^\circ$ with an average of 2.9$^\circ$ at $4 R_\odot$. In
eight events the angular widths of the rays at $3 R_\odot$ are
slightly larger than those at $4 R_\odot$. Note that the above
measurements of the ray width have uncertainties mainly
attributable to two factors. First, the projection effect makes
the measured width larger than the real value. Second, the
presence of nearby bright structures interferes with our
measurements of the ray profiles. Due to both factors our width
estimate could be an upper limit to the true value.

Comparing with the statistical study on post-CME rays by Webb
\textit{et al.} (2003) using the SMM data, we find three major
differences. The first is associated with the average lifetime or
duration of the rays. The duration of the rays we found varies
from 12 to 40 h with an average being 27 h, which is significantly
longer than that obtained in Webb \textit{et al.} (2003). The
second is related to the early morphological evolution of the
rays. We found that the rays grow gradually with increasing
brightness, while Webb \textit{et al.} (2003) concluded that the
ray structures appear suddenly with sudden brightening several
hours after the CME eruption. The third is that the ray width from
our measurements is generally larger than that given by Webb
\textit{et al.} (2003). These differences may be attributed to the
higher sensitivity and temporal resolution of LASCO than the SMM
coronagraph. Our results of the ray lifetime, appearance time and
angular width are in agreement with other studies using the LASCO
data (\textit{e.g.}, Vr\v snak \textit{et al.}, 2009). Note that
the events included in our study constitute a special group of
samples for the studies on the post-CME rays which are rich in
blobs. It remains elusive regarding the differences, if there are
any, between the blob-rich and blob-poor post-CME rays, which we
will not discuss further in this work.

In the following two subsections, we first analyze the event dated
on 21 September 2001 in detail to introduce our data analysis
method, and then give the results for the other ten events.

\subsection{Event on 21 September 2001}

In Figure 1(a) we show a white light image of the CME recorded by
LASCO C2 at 10:54 UT on 21 September 2001. The typical
three-component structure is clearly seen. The white circle in the
image represents the surface of the Sun, and the black one the
occulting disk of the C2 coronagraph. The horizontal and vertical
sizes of the image are both $6 R_{\odot}$. All the LASCO data used
in this study are processed using the standard routines included
in the SolarSoft package (http://www.lmsal.com/solarsoft/),
subtracting a background including the contribution of the F
corona and instrumental stray light. The CME first appeared in the
C2 FOV at 08:54 UT, and the bright ray structure was seen at 14:54
UT, \textit{i.e.}, 6 h later, with a length of about $0.6 R_\odot$
in the C2 FOV. After that, the ray started to grow gradually with
increasing brightness. At about 00:54 UT on the next day, the ray
started to diffuse gradually and became almost invisible in the C2
FOV at about 13:54 UT. The time duration of the ray is about 23 h
in total.

In Figure 1(b), we show the LASCO image with the ray structure
observed at 18:30 UT and indicated by a white arrow. The CPA of
the ray is 116$^{\circ}$ as measured at the bottom of the C2 FOV.
At this time, the ray was non-radial and tilted poleward by about
4$^\circ$, and the angular width of the ray at $3 R_{\odot}$ is
measured to be about 3.0$^\circ$ which corresponds to
$\approx$110\,000 km ($0.16 R_{\odot}$). The post-CME loops are
clearly seen from the EUV Imaging Telescope (EIT) 195 \AA\ images
shown in Figure 1(b). However, there was no flare that can be
associated with this event according to the CDAW database. As
explained previously, in this event the background X-ray emission
was as strong as the peak emission of a C3.0 flare. Therefore, it
is possible that the associated flare, if any, was too weak to be
observed. It is seen that the ray lay basically in between the
center of the CME ejecta and the top of the post-CME loops. To
observe the loops more clearly, in Figure 1(c) we show the
enlarged version of the EIT image inside the red square in Figure
1(b). It is measured that the altitude of the loop top is close to
$1.2 R_\odot$ from the solar center, which is a lower limit to the
real value considering the projection effect.

The electron density distribution around the ray structure can be
estimated using the $pB$ data which are obtained by LASCO C2 at
03:05 UT on 22 September and shown in Figure 1(d). The ray
structure is clearly identifiable. The $pB$ data along the four
arcs located at 3 (dot-dashed), 4 (dotted), 5 (dashed), and 6
(solid) $R_\odot$ are shown in Figure 1(e) to examine the
structure in a quantitative manner. The abscissa is the PA in the
range of 110$^\circ$ to 130$^\circ$, and the ordinate is the $pB$
value in units of the average brightness of the photosphere. We
see that the $pB$ intensity increases smoothly from the background
to the ray center up to $3 R_\odot$. Beyond $3 R_\odot$, the
increase of  $pB$ is more abrupt. The ratio of the local maximum
to the nearby background value of the $pB$ intensities, as read
from the $pB$ profiles at the vertical lines plotted in this
figure, increases from about 1.8 at $4 R_\odot$ to 2.1 (2.3) at 5
(6) $R_\odot$, respectively.

The blob structure flowing along the ray, as a moving density
enhancement, is best seen in the running difference images, which
are often used to highlight the brightness change in solar physics
studies and therefore employed for our blob study. In Figure 1(f),
we show an example of the blob observed in this event with the
difference image at 00:06 UT on 22 September. The blob, indicated
by the white arrow in the figure, is located at about $3.7
R_\odot$ revealing itself as the white-leading-black bipolar
structure seen in the image. The blob positions are measured from
the leading edge rather than from the center of the difference
structure, since it is easier to identify the former than the
latter in the difference images. Using both C2 and C3
observations, we obtain the height-time ($h-t$) plots as shown in
Figure 2(a) with plus signs, asterisks, and diamonds for the three
blobs observed in the event. The average velocities of the blobs
can be obtained by fitting the $h-t$ plots with straight lines,
which yield 462, 491, and 458 km s$^{-1}$, respectively. The
velocity variations with distance as well as the average
acceleration can be calculated with a second-order polynomial
fitting to all the $h-t$ data points. The fitted $h-t$ lines and
the calculated velocity profiles are plotted in Figures 2(a) and
2(b), respectively. In Figure 2(b), we also mark the corresponding
blob positions with the same symbols as those used in Figure 2(a).
We see that the first and third blobs showed significant
acceleration with average accelerations of 9.3 and 17.7 m
s$^{-2}$, respectively, while the second one moved with an almost
constant velocity. The number of blobs observed in the event, and
their average velocities and accelerations, as well as those for
the other ten events listed in Table 2 will be discussed in the
following subsection.

\subsection{The Other Events}

Using the same method, we have investigated the other ten events
listed in Table 1. To show the rays in question and the associated
observable post-CME loops, in Figure 3 we present the composites
of the LASCO C2 and EIT 195 \AA\ images with observational times
written on the corresponding panels. These images are chosen
because the concerned structures are clearly recognizable. The
post-CME rays are indicated by the white arrows in the panels. It
can be seen from the EIT images that post-CME loops are
discernible except in those given by panels (f) and (g) for the
two back-side events. Similar to the event shown in the above
subsection, it is clear that the rays were co-aligned with the
lines connecting the centroid of the ejecta to the corresponding
loop top, if present. It should be noted that the eruption in
event 5 involved a significant part of the lower solar atmosphere
including a simultaneous eruption of a long filament extending
from $\approx$S15E50 to $\approx$N09E42. The arch-like filament
moved outwards along the position angle in line with the observed
coronal ray. Underneath the erupting filament, growing bright
loops were observed that were located to the north-east of the
major loops producing the listed M2.8 flare. It is suggested that
this coronal ray was produced along with the erupted filament of
the complex eruption.

The altitudes of the loops vary in a range of 1.13 to 1.33
$R_\odot$ with an average of about 1.2 $R_\odot$. Again, these
values represent the lower limit to the real ones due to the
projection effect. The durations for these loop structures to be
observed are from twelve hours to a few days.

From Figure 3, we see that the brightness of the rays is different
from event to event. This is perhaps due to projection and/or
different intrinsic physical properties of rays in individual
events. The LASCO $pB$ data are available for four events.
Examining these data, we find that only for event 6 can the ray
structure be clearly discerned from the background and foreground
emissions, whose $pB$ data distribution, recorded at 21:00 UT on 9
January 2002, at distances of 3 (dot-dashed), 4 (dotted), 5
(dashed), and 6 (solid) $R_\odot$ have been superposed on Figure
1(e) with red lines. The range of the position angles is taken to
be from 70$^\circ$ to 90$^\circ$. It can be seen that the angular
width of the ray is about 2$^\circ$ -- 3$^\circ$ at this time, and
the ratio of the brightness maximum inside the ray to that around
is about 1.7 at 4 $R_\odot$, 1.8 at 5 $R_\odot$, and increases to
2.0 at 6 $R_\odot$. The $pB$ data used to calculate these ratios
have been marked in Figure 1(e) with short vertical lines. The
ratio becomes larger at distances higher than 3 $R_\odot$, similar
to what was observed for the other event. The ray is clearly
non-radial, tilted poleward by about 2$^\circ$ from 3 $R_\odot$ to
6 $R_\odot$ as seen from the $pB$ data. Comparing the $pB$ data
with the white light image shown in Figure 3(e), which was taken
earlier at 00:48 on 9 January, we find that the ray structure
moved by about 7$^\circ$ in the elapsed time between the two data
sets ($\approx$20 h).

The $h-t$ profiles of blobs in these events are obtained by
carefully examining the corresponding running difference images
from C2 and C3 observations, as introduced previously and shown in
Figure 4. The number of blobs observed in an event can be read
from this figure; the maximum number is 6 for the event on 18
November 2003. A total of 42 blobs were observed in all ten
events. From these observations, it seems that a major part of
these blobs are produced in a region from the top of the post-CME
loops, which lies at about 1.2 $R_\odot$, to the outer border of
the C2 FOV. The interval between the formation times of adjacent
blobs varies from 2 to 10 h.

With the linear and second-order polynomial fits to these $h-t$
profiles, we obtain the average velocities and accelerations. The
ranges of the two parameters have been included in Table 2. The
average velocity varies between 252--845 km s$^{-1}$ and the
average acceleration varies between -26.1--59.4 m s$^{-2}$. It is
found that the average velocity of blobs in an event does not
correlate well with the linear speed of the CME. Further detailed
studies on the correlation of the CME and the blob dynamics should
be conducted in the future with more events and projection effects
taken into account.

The velocity variations using the second-order fitting method for
the ten events are shown in Figure 5, from which we evaluate the
numbers of blobs experiencing acceleration or deceleration during
their propagation. It is found that, including the event shown in
Figures 1 and 2, there are 26 (9) blobs that were accelerated
(decelerated) after their first observations by C2, and 10 blobs
whose velocities did not change appreciably with the total
velocity variation below 60 km s$^{-1}$ and the average
acceleration less than 3 m s$^{-2}$. Physical implications of this
statistical result will be further discussed in the following
section. It is also found that the difference in dynamical
properties of blobs in a specific event can be rather large. For
instance, for the five blobs observed in event 10 on 4 November
2003, the maximum (minimum) average velocity was 786 (586) km
s$^{-1}$ which corresponded to the first (third) blob, and the
average accelerations were in the range of -26.1 m s$^{-2}$
(minimum) to 22.9 m s$^{-2}$ (maximum). The velocities and
accelerations in one event seem to have no strong relationship
with their occurrence order.

It should be pointed out that the blob events 6, 10, and 11 listed
in Table 2 have also been respectively studied by Ko \textit{et
al.} (2003), Ciaravella and Raymond (2008), and Lin \textit{et
al.} (2005). The dynamical parameters of blobs provided in these
previous studies are slightly different from those given in the
present study. This is mainly due to two reasons. The first is
related to the subjectivity in determining the position of the
blobs. The second is due to the different altitude ranges used to
measure the $h-t$ profiles which affected the results of the
fitting. Nevertheless, these factors do not change the major
results of our statistical study.

\section{Discussion on Blob Dynamics and Associated Magnetic Reconnection}

In Figure 6, we display the velocity-distance profiles of all 45
CME-ray-blobs (black circles) studied in this work together with
those of 106 streamer blobs (red pluses) observed in 2007 and
reported by Song \textit{et al.} (2009). This allows us to compare
the dynamics of the two groups of blobs statistically. We see that
the most obvious differences between the two data sets lies in the
range of velocities. For instance, at 4 $R_\odot$ the range is 300
-- 800 km s$^{-1}$ with an average of 450 km s$^{-1}$ for the
former group, and 50 -- 250 km s$^{-1}$ with an average of 150 km
s$^{-1}$ for the latter one. At 8 (12) $R_\odot$, the velocity
range is 330 -- 800 (330 -- 750) km s$^{-1}$ with an average of
565 (540) km s$^{-1}$ for the former, and 150 -- 320 (200 -- 400)
km s$^{-1}$ with an average of 230 (300) km s$^{-1}$ for the
latter. At a fixed distance, the velocity of the CME blobs shows a
range as large as 400 -- 800 km s$^{-1}$, which is much more
significant than that of the streamer blobs ($\leq$ 300 km
s$^{-1}$). Therefore, it can be summarized that the velocity
values of the CME blobs, as well as their velocity variations, are
in general much larger than those of the streamer blobs. It is
interesting to find that the scatter plot between the velocities
of the two groups can be well separated by the blue line plotted
in Figure 6, which is obtained using the following method. First,
we fit the velocity data of both groups with two quadratic curves
$v_{1}=A_{1}r^{2}+B_{1}r+C_{1}$ and
$v_{2}=A_{2}r^{2}+B_{2}r+C_{2}$. We then average the parameters to
obtain $A=(A_{1}+A_{2})/2$, $B=(B_{1}+B_{2})/2$ and
$C=(C_{1}+C_{2})/2$ and construct the blue line as
$v=Ar^{2}+Br+C$. By doing so, we have obtained
$v=-0.1r^{2}+13.7r+246.9$ with $v$ and $r$ in units of km s$^{-1}$
and $R_\odot$, respectively. It is found that about 83\% of the
CME-blob measurements are above the line while almost all (98\%)
of the streamer-blob measurements are below it.

Another obvious difference between the dynamics of CME blobs and
streamer blobs can be revealed from Figures 6 and 5, and previous
studies on the streamer blobs (\textit{e.g.}, Wang \textit{et
al.}, 2000; Song \textit{et al.}, 2009). That is, most if not all
streamer blobs present a gradual acceleration, while in the case
of CME blobs the situation is more complex with 58\% (20\%) being
accelerated (decelerated) and 22\% moving with nearly constant
velocities.

Stimulated by the above two differences, in the following we will
discuss possible physical factors that may affect the formation
and acceleration of the two types of blobs. Physically, the blob
dynamics are determined mainly by their formation process, which
is believed to be magnetic reconnection at the current sheet
giving rise to the initial acceleration of blobs, as well as the
interaction between the background plasma and the magnetic fields
during the blob propagation. It is known that the CME blobs are
produced along the rays in the aftermath of a CME, while the
streamer blobs are released from atop of a helmet streamer along
the bright plasma sheet. The magnetic field strength along the
post-CME rays has been estimated by Ko \textit{et al.} (2003) to
be about 0.69 G at about 3 $R_\odot$. Vr\v snak \textit{et al.}
(2009) obtained the field strength of 1.7 G at 1.5 $R_\odot$,
which means 0.43 G at 3 $R_\odot$ assuming an $r^{-2}$ dependence.
These field strengths are appreciably stronger than those along
the streamer plasma sheet, which are around 0.1 G at 3 $R_\odot$
according to a seismological study of a wave phenomenon in the
streamer observed with LASCO (Chen \textit{et al.}, 2010, 2011;
Feng \textit{et al.}, 2011). Consequently, the associated magnetic
reconnection may have different impacts on the initial blob speeds
which are significantly larger in the CME blobs than in the
streamer ones.

The unanimous gradual acceleration of streamer blobs has been
regarded as evidence indicating the important roles played by the
surrounding accelerating solar wind on their dynamics, and
therefore they have been taken as velocity tracers of the wind
along the plasma sheet (Sheeley \textit{et al.}, 1997; Song
\textit{et al.}, 2009). On the other hand, CME blobs discussed
here show acceleration, deceleration, or negligible velocity
change at all. This may be explained by the large differences of
flow and magnetic field conditions in individual events with
distinctive consequences of the coupling process between the blobs
and their surroundings.

The acceleration in more than half of the CME-blob events may be
attributed to the tension force associated with the magnetic field
lines reconnected below the blobs. These lines are connected to
and dragged outwards by the ejecta with large concave outward
curvature, and capable of producing persistent acceleration of the
blobs by the sling-shot effect.

In summary, the differences between the magnetic reconnection
processes leading to CME blobs and streamer blobs are threefold.
(1) The dynamical behavior of their products, \textit{i.e.},
blobs, including their velocity ranges and
acceleration/deceleration characteristics, are totally different
from each other as already presented above in detail. (2) The
associated magnetic structures and field strengths are
considerably different. As mentioned already, the magnetic
reconnection leading to streamer blobs is related to the streamer
cusp and heliospheric plasma/current sheet with weaker field
strength, while that leading to CME blobs is related to the
post-CME ray structure hosting the CME current sheet with
relatively stronger field strength. (3) The triggering process for
the reconnection may be different. The one producing streamer
blobs is possibly driven by fluid instabilities (Chen \textit{et
al.}, 2009) and the one producing CME blobs is mostly excited by
the tearing mode instability (\textit{e.g.}, Ko \textit{et al.},
2003; Lin \textit{et al.}, 2009).

Finally, we note that the post-CME ray structures, the blobs
flowing along them, as well as the associated magnetic
reconnection processes, should be further investigated with
multi-viewpoint coronagraphs onboard the STEREO spacecraft (Kaiser
\textit{et al.}, 2008) together with SOHO/LASCO to take into
account the projection effect of the motion and their
three-dimensional features.

\section{Summary}

A survey on post-CME rays and blobs has been carried out using the
LASCO data from 1996 to 2009. Eleven events were found with
clearly-observable post-CME rays along which multi-blobs flow
outwards. The post-CME rays were suggested to be associated with
the current sheets formed as an aftermath of the CME eruptions,
and the blobs are explained as products of magnetic reconnection
along these current sheets. The ray morphology and blob dynamics
were investigated in these eleven events in a statistical manner.
It was found that the angular widths of the post-CME rays at a
given time vary from 2.1$^\circ$ -- 6.6$^\circ$ (2.0$^\circ$ --
4.4$^\circ$) with an average of 3.5$^{\circ}$ (2.9$^{\circ}$) at 3
$R_\odot$ (4 $R_\odot$), the duration of events are from 10 h to a
few days with an average of 27 h, and most rays are aligned
non-radial while a few of them exhibit apparent changes in the
position angle. In all the 45 blobs included in the study, more
than half of them (26) exhibit acceleration, nearly a quarter of
them (9) show deceleration while the rest of them (10) keep their
velocity almost unchanged during outward propagation. Comparing
the velocity profiles of these post-CME blobs with those that are
released from atop of helmet streamers, we found that the velocity
values and variation ranges for the CME blobs are significantly
larger than those of the streamer blobs.

The dynamical behaviors of the two groups of blobs suggest that
the underlying physics of associated magnetic reconnection are
different for these two groups. The differences are threefold.
First, the reconnection processes have different observational
manifestations as revealed by the blob dynamics mentioned above.
Second, the reconnection processes are associated with different
magnetic structures being either the post-CME rays hosting current
sheets or streamer cusp region with a much weaker magnetic field
strength. Last, their triggering processes may be different with
the post-CME reconnection being excited mostly by tearing mode
instabilities and the streamer-cusp reconnection being driven
possibly by ideal fluid instabilities.

\begin{acks}
The LASCO data used here are produced by a consortium of the Naval
Research Laboratory (USA), Max-Planck-Institut f\"ur Aeronomie
(Germany), Laboratoire d'Astronomie (France), and the University
of Birmingham (UK). SOHO is a project of international cooperation
between ESA and NASA. We thank Dr. A. Vourlidas for helping us
analyze the LASCO $pB$ data. This work was supported by grants
NNSFC 41028004, 40825014, 40890162, Natural Science Foundation of
Shandong Province ZR2010DQ016, Independent Innovation Foundation
of Shandong University 2010ZRYB001, and the Specialized Research
Fund for State Key Laboratories in China. Y. Chen is also
supported by A Foundation for the Author of National Excellent
Doctoral Dissertation of PR China (2007B24). B. Li is supported by
the grant NNSFC 40904047, and L.D. Xia by 40974105.
\end{acks}



\begin{figure}[tphb]
\includegraphics[width=1\textwidth]{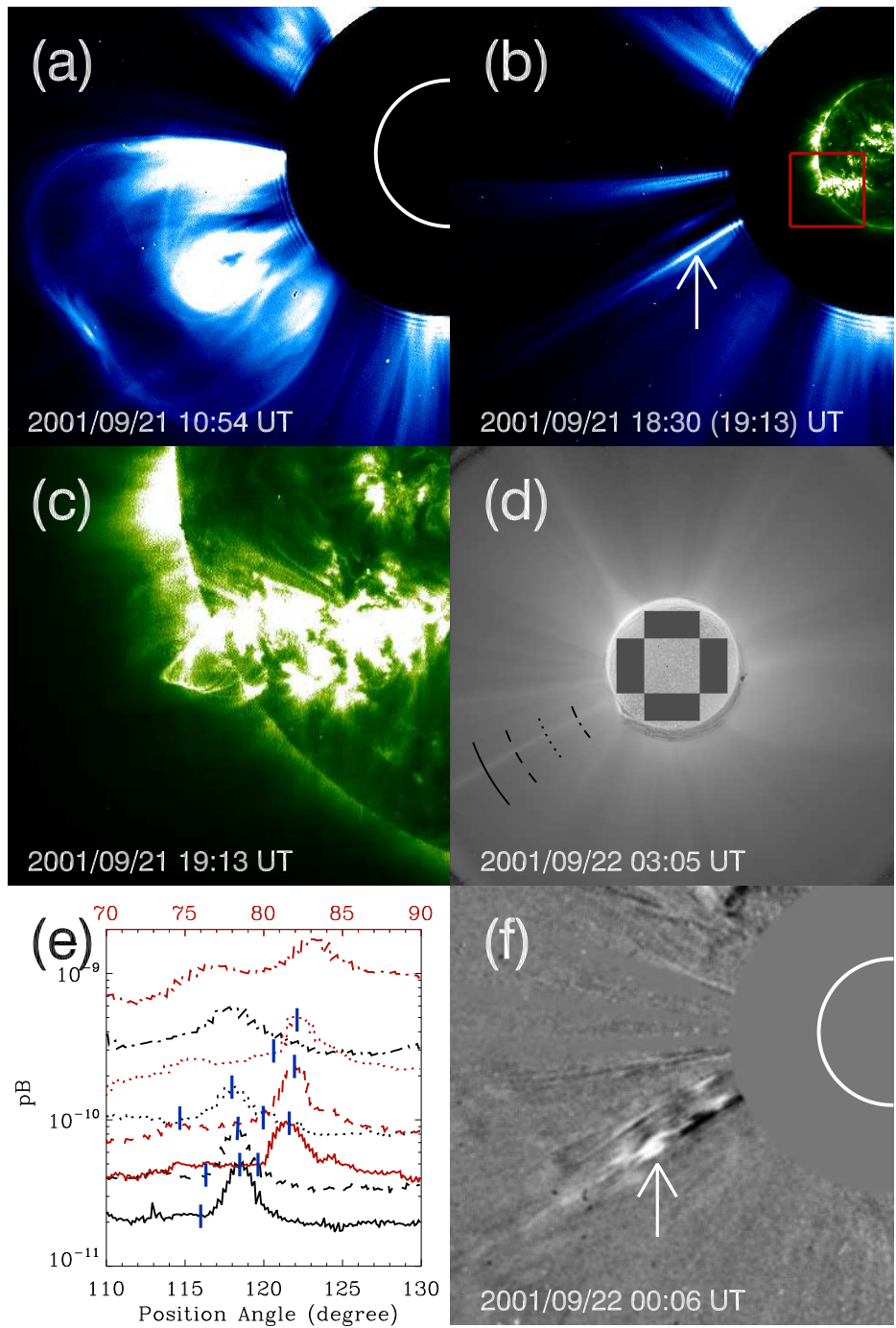}
\caption{(a) The white-light image observed by LASCO C2 at 10:54
UT for the event on 21 September 2001. (b) Composite image from
LASCO C2 (18:30 UT) and EIT 195 \AA\ (19:31 UT) showing the
post-CME ray, as denoted by the white arrow, and the post-CME
loops. (c) The enlarged version of the EIT image inside the red
square in Figure 1(b). (d) pB data recorded by LASCO at 03:05 UT
on 22 September 2001. The four arcs are at 3, 4, 5, and 6
R$_\odot$, along which the pB intensities are plotted in the next
panel. (e) pB intensities in units of median solar brightness
versus position angles at 3 (dot-dashed), 4 (dotted), 5 (dashed),
and 6 (solid) R$_\odot$ for the 2001/09/21 event (black lines) and
the 2002/01/08 event (red lines). (f) Running difference image at
00:06 on 22 September 2001 showing the presence of a blob
structure. [See the electronic edition for a color version of this
figure.]}
\end{figure}

\begin{figure}[tphb]
\includegraphics[width=1\textwidth]{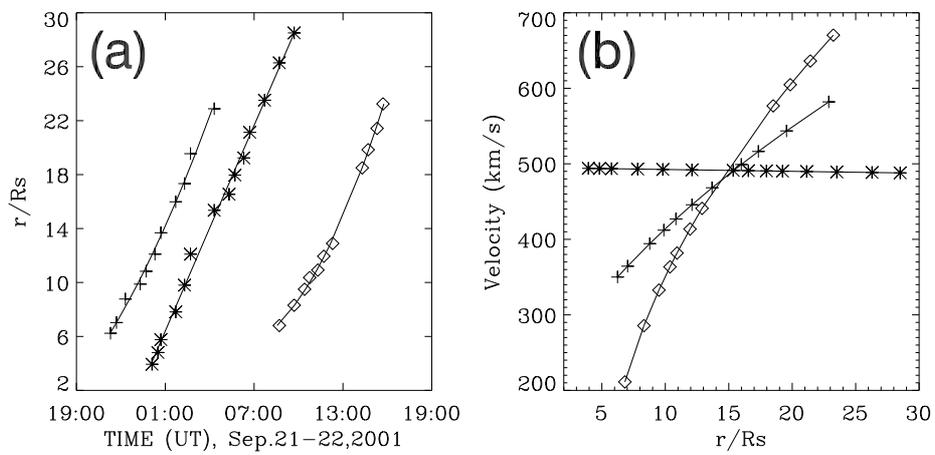}
\caption{(a) Height-time profiles of the three blobs measured in
the event on 21 September 2001 and represented by plus signs,
stars, and diamonds. The lines are given by the second-order
polynomial fitting to all the h-t data points. (b) Radial
variations of the blob velocities deduced from the h-t fittings.
The corresponding positions for the blobs are marked with the same
symbols as that used in Figure 2(a).}
\end{figure}

\begin{figure}[tphb]
\includegraphics[width=0.6\textwidth]{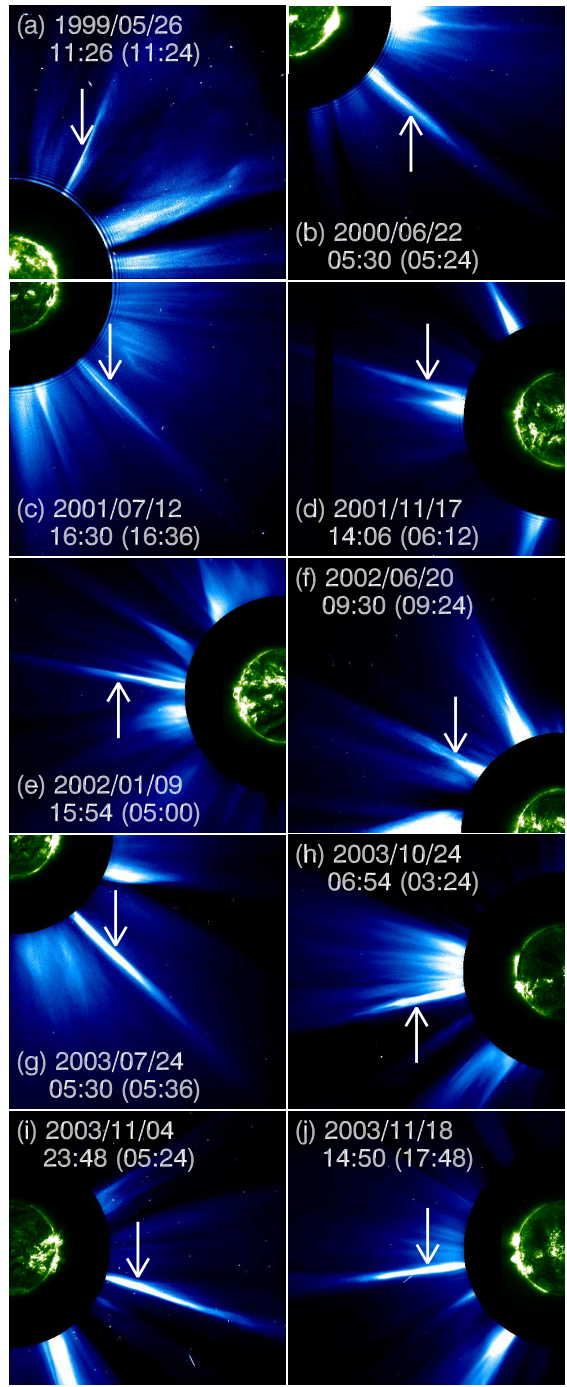}
\caption{Composite images from LASCO C2 and EIT 195 \AA\ showing
post-CME rays, denoted by the white arrows, and post-CME loops for
the ten events. The observation time for LASCO is given in each
panel, the time in parenthesis corresponds to the EIT data. [See
the electronic edition for a color version of this figure.]}
\end{figure}

\begin{figure}[tphb]
\includegraphics[width=1\textwidth]{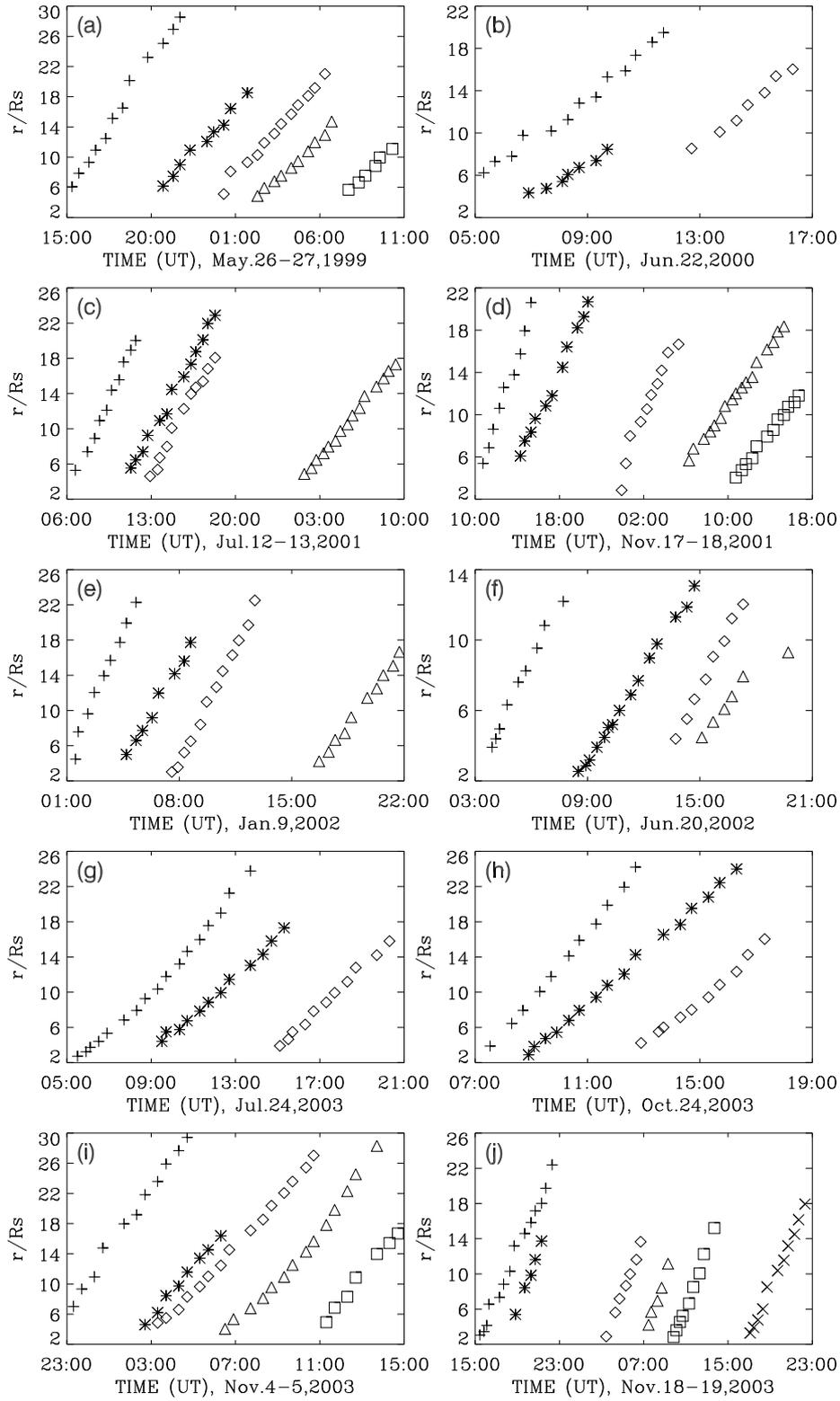}
\caption{Height-time profiles of blobs measured in the ten events
shown in Figure 3. The blob occurrence order is indicated by
various symbols.}
\end{figure}

\begin{figure}[tphb]
\includegraphics[width=1\textwidth]{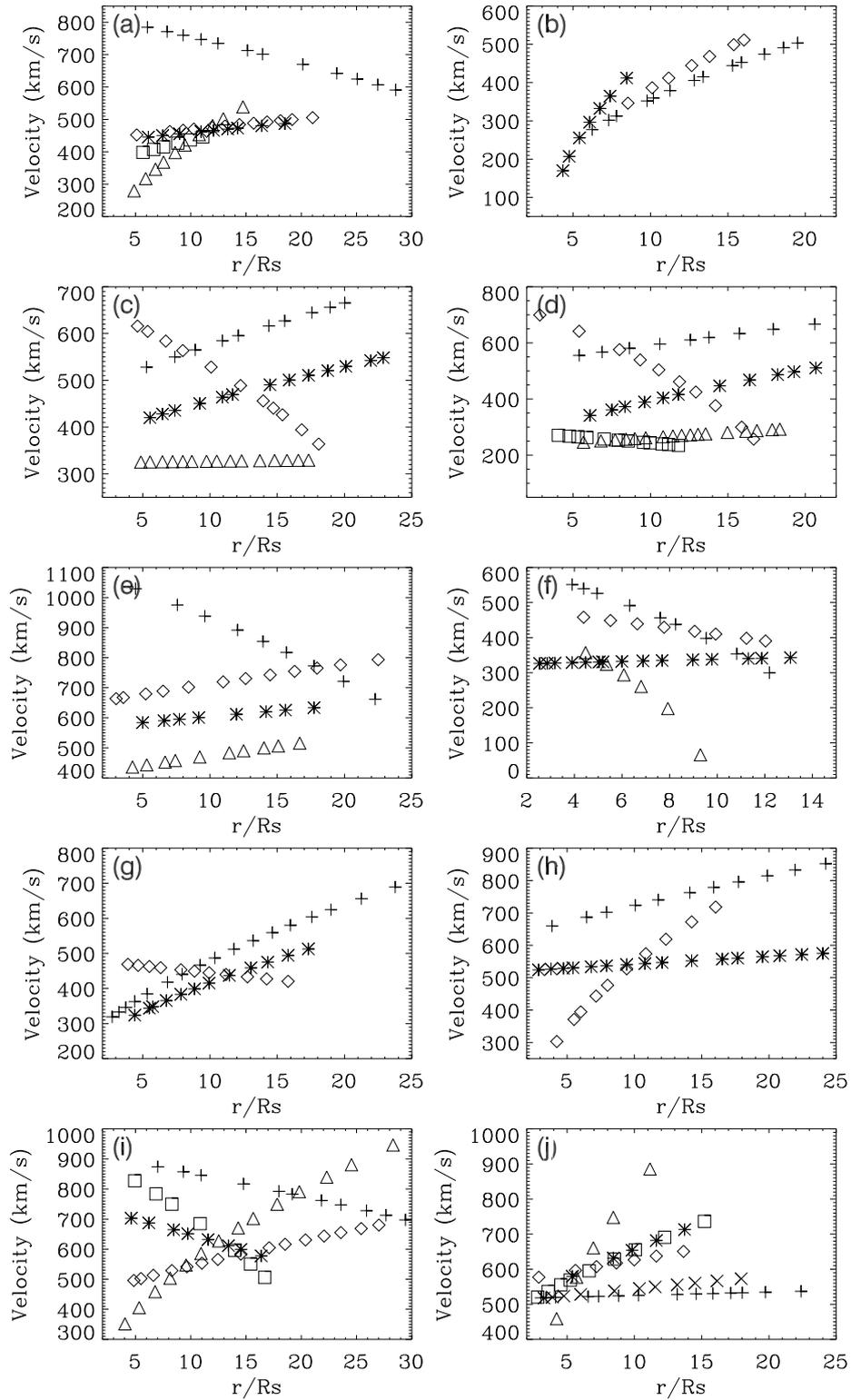}
\caption{The fitted velocities of blobs as a function of
heliocentric distance for the ten events shown in Figure 3. The
blob occurrence order is indicated by the same symbols used in
Figure 4.}
\end{figure}

\begin{figure}[tphb]
\includegraphics[width=1\textwidth]{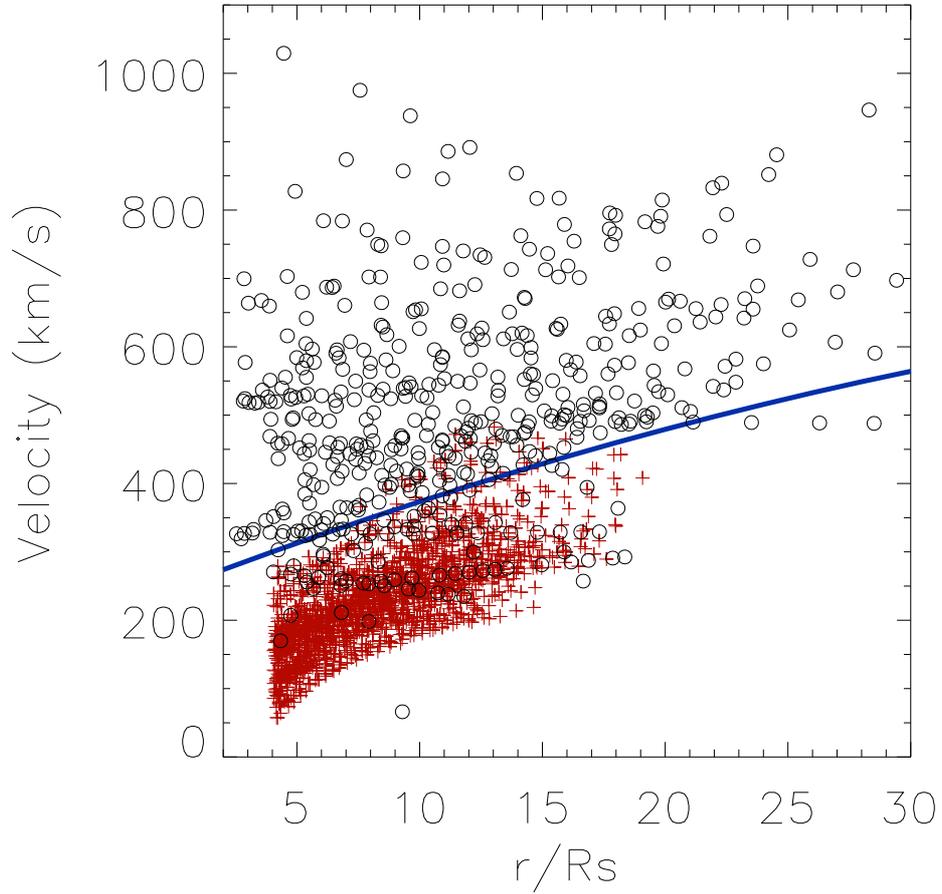}
\caption{Velocity-distance profiles for the 45 post-CME blobs
collected in this study (black circles) and the 106 streamer-blobs
(red pluses) given by Song \textit{et al.} (2009). (The blue line
is drawn to separate their velocity distributions.) The blue line,
serving as a good separator of the two data sets, is given by the
average of the polynomial fittings to the two groups of velocity
measurements. [See the electronic edition for a color version of
this figure.]}
\end{figure}

\end{article}

\end{document}